\documentclass[aps,prl,twocolumn,groupedaddress]{revtex4-2}
\usepackage{amssymb,amsthm}
\usepackage{amsmath,amssymb,amsfonts,bm,amscd}
\usepackage{graphicx}

\usepackage{caption, subcaption}
\usepackage{multirow}
\usepackage{tensor}

\usepackage{tikz}

\usepackage{hyperref}

\newtheorem{conj}{Conjecture}
\newtheorem{cor}{Corollary}

\def\fkg{\mathfrak{g}}

\def\fsu{\mathfrak{su}}
\def\fso{\mathfrak{so}}

\def\CM{\mathcal{M}}
\def\CN{\mathcal{N}}

\def\CT{\mathcal{T}}

\def\bbP{\mathbb{P}}

\def\bbZ{\mathbb{Z}}
\def\bbV{\mathbb{V}}

\def\CM{{\cal M}}
\def\CN{{\cal N}}

\def\beq#1\eeq{\begin{align}#1\end{align}}

\usepackage{enumitem}

\def\fsu{\mathfrak{su}}
\def\fso{\mathfrak{so}}

\def\bbZ{\mathbb{Z}}

\def\beq#1\eeq{\begin{align}#1\end{align}}
\theoremstyle{definition}

\begin{document}

%\preprint{}

\title{Mirror symmetry for circle compactified 4d \texorpdfstring{$A_1$}{A1} \texorpdfstring{class-$\mathcal{S}$}{class-S} theories}


\author{Yiwen Pan}
%\thanks{} %\altaffiliation{}
\affiliation{Department of Physics,  Sun Yat-Sen University, Guangzhou, Guangdong, China}

\author{Wenbin Yan}
%\thanks{}
\affiliation{Yau Mathematics Science Center, Tsinghua University, Beijing, China}


\date{\today}

\begin{abstract}
In this letter, we propose a 4d mirror symmetry for the class-$\mathcal{S}$ theories which relates the representation theory of the chiral quantization of the Higgs branch and the geometry of the Coulomb branch. We study the representation theory by using the 4d/VOA correspondence, (defect) Schur indices and (flavor) modular differential equations, and match the data with the fixed manifolds of the Hitchin moduli spaces. This correspondence extends the connection between Higgs and Coulomb branch of Argyres-Douglas theories, and can provide systematic guidance for the study of the representation theory of vertex operator algebras by exploiting  results from Hitchin systems.
\end{abstract}

%\keywords{}

\maketitle

\section{Introduction}

The 3d mirror symmetry \cite{Intriligator:1996ex} is a remarkable duality between a pair of 3d  theories with 8 supercharges such that the Coulomb branch (CB) of one  is the Higgs branch (HB) of the other one and vice versa. It is an invaluable tool to study moduli spaces of 3d theories, and also deeply related to 3d symplectic duality \cite{braden2012quantizations,braden2014quantizations} in mathematics.
One may also ask whether similar mirror symmetry exists for 4d $\CN=2$ theories with also 8 supercharges. At first glance this seems impossible, as the HBs and CBs in 4d have different geometries, and can never exchange roles like in 3d. However, inspired by the 3d symplectic duality where the quantization of 3d HB is connected to the cohomology of 3d CB, one may view the 4d mirror symmetry as the relation between certain ``quantization'' of HB and the geometry of CB of a given 4d $\CN=2$ theory. A natural choice is the vertex operator algebra (VOA) from the 4d/VOA correspondence \cite{Beem:2013sza} which is viewed as the chiral quantization of the Higgs branch \cite{Beem:2017ooy, Song:2017oew,Arakawa:2018egx}. Such 4d mirror symmetry was proposed and proved for 4d $\CN=2$ generalized Argyres-Douglas (AD) theories \cite{Shan:2023xtw,Shan:2024yas} (see also \cite{Fredrickson:2017yka, Fredrickson:2017jcf}).

In this letter we provide strong evidences for the correspondence between the chiral quantization of HB and geometry of CB for a large class of 4d $\CN=2$ SCFTs: the $A_1$ class-$\mathcal{S}$ theories $\mathcal{T}_{g, n}$ constructed by compactification of 6d $(2,0)$ SCFTs on a genuse $g$ Riemann surface $\Sigma_{g,n}$  with $n$ regular punctures \cite{Gaiotto:2009we,Gaiotto:2009hg}. The flavor symmetry of $\CT_{g,n}$ is $\fsu(2)^{\otimes n}$ which may get enhanced in some cases. The HB of $\CT_{g,n}$ is the Moore-Tachikawa variety \cite{Moore:2011}, while the CB of $\CT_{g,n}$ compactified on a circle of finite radius is isomorphic to the moduli space of Hitchin system on $\Sigma_{g,n}$ \cite{Gaiotto:2009hg}.

The chiral quantization $V_{g,n}$ of the HB of $\CT_{g,n}$  has been constructed in \cite{Arakawa:2018egx}. It is expected that $V_{g,n}$ has a finite number of simple modules (SM). Characters of theses SMs, as meromorphic functions of the flavor fugacities and modular parameter $q$, satisfy a set of partial differential equations called the flavor modular differential equations (FMDEs) \cite{Zheng:2022zkm}. The quasi-modularity \cite{Pan:2023jjw} of FMDEs implies that  characters belong to a representation $\bbV^\text{mod}_{g,n}$ of the modular group. Usually the dimension of $\bbV^\text{mod}_{g,n}$ is larger than the number of SMs, where the extra solutions may be identified as characters of logarithmic modules. Ordinary modules (OM) are special SMs whose characters are still convergent when setting all flavor fugacities to $1$. Unflavored characters of OMs satisfy a modular differential equation (MDE) \cite{Arakawa:2016hkg,Beem:2017ooy} and belong to a smaller representation $\bbV^\text{ord}_{g,n}$ of the modular group. Practically, we often use the correspondence between the 4d defects and VOA modules \cite{Beem:2013sza, Cordova:2017mhb,Cordova:2016uwk,Pan:2021mrw,Guo:2023mkn} to obtain characters of SMs of $V_{g,n}$ \cite{Pan:2021mrw,Guo:2023mkn,Pan:2024bne}, and extract data of highest weight state (HWS) from characters. The basis of $\bbV^\text{mod}_{g,n}$ or $\bbV^\text{ord}_{g,n}$ can be constructed using modular transformations \cite{Pan:2024bne}.

The geometry of the CB of $\CT_{g,n}$ is described by the moduli space $\CM_{g,n}$ of the $SU(2)/\bbZ_2$ Hitchin system on $\Sigma_{g,n}$. We are interested in the set of fixed manifolds (FM) denoted by $\CM_{g,n}^T$ of $\CM_{g,n}$ under the $U(1)_r$ from the superconformal algebra. $\CM_{g,n}^T$'s  are studied in \cite{hitchin1987self} for $n=0$  and in \cite{boden1996moduli,nasatyr1995orbifold} for $n>0$. They provide a description for each FM and the critical value of the moment map of $U(1)_r$ on each FM, which will be crucial in matching with data from $V_{g,n}$. Notice that $\CM_{g,n}$ is also homeomorphic to the character variety $C_{g,n}$ on $\Sigma_{g,n}$ whose geometries are studied in \cite{hausel2008mixed}.

Using the above data from both HB and CB, we observe a bijection between the SMs of $V_{g,n}$ and FMs of $\CM_{g,n}$ of the class-S theory $\CT_{g,n}$. The HWS of each SM is completely determined by the value of moment map of the corresponding FM. Moreover, Jordan type of the modular matrices on $\bbV^\text{mod}_{g,n}$ are determined by dimensions of each FM and $g$. The later observation also provides an interpretation of logarithmic modules as certain elements in the cohomology of $\CM_{g,n}$. Our results show that there is a deep connection between the chiral quantization of the HB and geometry of CB for a large class of 4d $\CN=2$ theories, suggesting a correspondence between representation theory of VOAs and geometry of Hitchin systems in more general terms. 
%It would be interesting to explore the potential underlying mechanism uniting the two both physically and mathematically.
In this letter we will provide statements of our proposal and crucial evidences and leave details in an accompany paper \cite{Pan:WIP}.

\section{Main results}
\label{sec:main}

%Here we summarize our conjectures on the simple modules, ordinary modules of $V[\mathcal{T}^\mathfrak{g}_{g,n}]$, their modular property, and their relation with the structure of the fixed manifolds of $\mathcal{M}[\mathcal{T}_{g,n}^\mathfrak{g}]$. In the subsequent discussions we will provide evidences to support these conjectures with series of examples. 

First consider the case when $n=0$. As there is no flavor symmetry from punctures, the HWS of $L_a$ is determined by its conformal dimension $h(L_a)$ \footnote{There is an accidental flavor symmetry discussed in \cite{Beem:2013sza}, however, we will not discuss this accidental flavor symmetry in this letter.}, and all $L_a$'s are OMs. We have
%\YP{[I changed $i$ to $a$ to avoid confusion with simple root label $i$]} 
\begin{conj}\label{conjn0}When $n=0$, there is a bijection between the $g$ simple modules $\{L_a\}$ of $V_{g,0}$ and the $g$ fixed manifolds $\CM^T_{g,0}\equiv\{M_a\}_{0\leq a\leq g-1}$ of $\CM_{g,0}$ such that
\begin{equation}
\label{eq:bijectionn0}
h(L_a)=\mu(M_a) -g+\frac{3}{2}-\frac{1}{2}\delta_{\mu(M_a),0}.
\end{equation}
Moreover, the Jordan type of modular $T$ matrix is 
\begin{equation}
[(1-\delta_{\mu(M_a),0})\dim M_a +g+1]_{M_a\in \CM_{g,0}^T},
\end{equation}
with $g$ Jordan blocks.
\end{conj}
The order of MDE is just the sum of sizes of all Jordan blocks of $T$. Our results compute the order of MDE in terms of dimensions of FMs.

When adding regular punctures on the Riemann surface, one needs also assign a parabolic parameter $\alpha_i$ for the $i$-th puncture. The value of moment map $\mu$ on each FM will be a function of  parameters $\{\alpha_i\}_{0\leq i\leq n}$. The correspondence in this case is given by the following.
\begin{conj}
\label{conj2}
When $n>0$ and assuming $0<\alpha_n<\cdots<\alpha_1<1/2$, there is a bijection between the simple modules $\{L_a\}$ of $V_{g,n}$ and the fixed manifolds $\CM^T_{g,n} = \{M_a\}$ such that for even $n$,
\begin{eqnarray}
\label{eq:bijneven}\mu_{M_a} - \mu_\text{max}|_{\alpha_i\mapsto -\omega^{(i)}}
= -h(L_a) +\sum_{i=1}^n \lambda_i \omega^{(i)},
\end{eqnarray} 
and for odd $n$,
\begin{equation}
  \label{eq:bijnodd} \mu_{M_a} - \mu_\text{max}|_{\alpha_i\mapsto \frac{1}{2}+\omega^{(i)}}
  = -h(L_a) +\sum_{i=1}^n \lambda_i \omega^{(i)}
\end{equation}
Here $\mu_{max}$ is the maximum of all critical values of the moment map, $\omega^{(i)}$ is the fundamental weight of the $i$-th flavor $\fsu(2)$. The HWS of $L_a$ has conformal dimension $h(L_a)$ and transforms in the representation $[\lambda_1,\lambda_2,\cdots,\lambda_n]$ under the flavor symmetry $\fsu(2)^{(1)}\times\cdots \fsu(2)^{(n)}$. Moreover, the Jordan type of modular $T$ (resp. $STS$) is 
\begin{equation}
\label{eq:dimJord}
[(1-\delta_{\mu(M_a),0})\dim M +g+1]_{M_a\in \CM_{g,n}^T},
\end{equation}
for $n$ even (resp. odd).
\end{conj}
When $g=0$ the total number of simple and logarithmic modules is the same as the dimension of the cohomology of $\CM_{0,n}$, hinting strongly that logarithmic modules correspond to elements in the cohomology ring of $\CM_{0,n}$.

As the moment map of an FM completely determines the HWS of the corresponding SM, we can identify all the OMs, and compute the modular matrices among them (and some of the logarithmic modules) such as $T^\text{ord}$, $(STS)^\text{ord}$.
\begin{cor}
  \label{corollary1}All OMs of $V_{g,n}$ correspond to FMs in the subset $\CM^\text{ord}_{g,n}\equiv\{M_{d,(1,1,\cdots,1)}\}_{-n+\sum_i\alpha_i<d\leq g-1-n/2}$ (resp. $\CM^\text{ord}_{g,n}\equiv\{M_{d,(0,1,\cdots,1)}\}_{-n+1-\alpha_1+\sum_{i>1}\alpha_i<d\leq g-1-(n-1)/2}$) of $\CM^T_{g,n}$ when $n$ is even (resp. odd) such that
\begin{equation}
  h(L^\text{ord}) = -(\mu(M)-\mu_{max}).
\end{equation}
The Jordan type of $T^\text{ord}$ (resp. $(STS)^\text{ord}$) is $[\dim M+g+2]_{M\in \CM^\text{ord}_{g,n}}$ (resp.  $[\dim M+g+1]_{M \in \CM^\text{ord}_{g,n}}$).
\end{cor}
Again, the total size of $T^\text{ord}$ or $(STS)^\text{ord}$ (hence the order of MDE) can be computed using the dimensions of FMs corresponding to OMs.

As mentioned before $\CM_{g,n}$ is homeomorphic to the character variety $C_{g,n}$. We also find the following  relations among $C_{g,n}$, $\CM_{g,n}$  and $V_{g,n}$.
\begin{conj}\label{conj:chavar}Let $PH_{g,n}(q)$ be the pure part of the mixed Hodge polynomial of $C_{g,n}$ ( \cite[Conjecture 1.2.1(iv)]{Hausel_2011} ). Define the  renormalized polynomial (RPMHP) $P_{g,n}(q)$
\begin{equation}
P_{g,n}(q)\equiv q^{\delta_{n,0}+g(1-\delta_{n,0})+\delta_{g,0}-\frac{d_{g,n}}{2}} PH_{g,n}(q) = \sum_{i}a_i q^{d_i} \nonumber \ .
\end{equation}
The total number of FMs of $\CM_{g,n}$ (i.e. SMs of $V_{g,n}$ is $P_{g,n}(1)=\sum_i a_i$, $\dim \bbV^{mod}_{g,n} =\frac{dP_{g,n}(q)}{dq}|_{q=1}= \sum_i a_i d_i$, and the Jordan type of the modular matrix is $[d_i^{a_i}]$. Here $d_{g,n}=-(2-2g-n)\dim\fkg$ is the dimension of $C_{g,n}$. 
\end{conj}
This conjecture also builds a bridge between RPMHP of character varieties and FMs of Hitchin moduli spaces. To the best of our knowledge, this correspondence was not noticed in math literature.

\section{Examples}
\label{sec:examples}

In this section we provide  evidences for our proposal.

\subsection{\texorpdfstring{$g\geq2, n=0$ cases}{}}

%We begin with the simplest class-$\mathcal{S}$ theories with $n = 0$ puncture. Such a theory does not have any $SU(2)$ flavor symmetry associated to punctures. Therefore, we can focus on the ordinary modules and log modules \footnote{There is an accidental flavor symmetry discussed in \cite{Beem:2013sza}, however, we will not discuss this accidental flavor symmetry in this paper.}, whose characters are solutions to the corresponding unflavored MLDE.

First consider $g=2$. The VOA $V_{2,0}$ has central charge $c = -26$ and has been studied \cite{Kiyoshige:2020uqz,Beem:2021jnm}. The corresponding MDE is  \cite{Beem:2017ooy},
{\small\begin{align}
 \Big[D_q^{(6)} - & \ 305 E_4 D_q^{(4)} - 4060E_6 D_q^{(3)}
      + 20275E_4^2 D_q^{(2)} \nonumber \\
      & \ + 2100E_4 E_6 D_q^{(1)}  - 68600(E_6^2 - 49125E_4^3) \Big]\operatorname{ch} =0 . \nonumber
\end{align}}
There are two non-logarithmic solutions which correspond to characters of OMs of $h=0$ and $-1/2$, together with four logarithmic solutions. % to four log modules. 
%There are two non-logarithmic (non-log) solutions
%\begin{equation}
%   \frac{1}{2}\eta(\tau)^2(E_2 + \frac{1}{12}), \qquad
%  \eta(\tau)^2 \ ,
%\end{equation}
%which correspond to the vacuum module ($h=0$) and another module with $h=-1/2$ respectively. There are also four logarithmic (log) solutions given by
%\begin{align}
%  \operatorname{ch}^{\text{long}}_1 = & \ \eta(\tau)^2(- \frac{1}{24} i \tau  + \frac{1}{4\pi} \tau^2  - \frac{i}{2}\tau^3  E_2(\tau) )\nonumber\\
%  \operatorname{ch}^{\text{long}}_2 = & \ \eta(\tau)^2(- \frac{1}{24} i \tau  + \frac{1}{4\pi}  + \frac{1}{2\pi} \tau  - \frac{i}{2}  E_2(\tau)) \nonumber\\
%  & \  - \frac{3}{2}i \tau \eta(\tau)^2 E_2(\tau) - \frac{3}{2}i \tau^2 \eta(\tau)^2 E_2(\tau) \\
%  \operatorname{ch}^{\text{long}}_3 = & \ + \eta(\tau)^2(\frac{1}{2\pi} - 3i  E_2 - 3i\tau E_2(\tau))\nonumber\\
%  \operatorname{ch}^{\text{long}}_4 = & \ \eta(\tau)^2(\frac{i}{\pi}\tau  - \frac{3i}{2\pi} \tau^2  + 3 \tau^2  E_2(\tau) - 3 \tau^3  E_2(\tau)) \ . \nonumber
%\end{align}
They span a six dimensional representation of the modular group $SL(2,\bbZ)$. In particular, the Jordan type of its $T$-matrix is $[4,2]$. On the other hand, the set of FMs $\CM^T_{2,0}$ has two elements $M_0, M_1$ with $\mu=0,1/2$ and dimension $\{2, 1\}$,  giving conformal dimension $\{-1,0\}$ using \eqref{eq:bijectionn0}. Also $[\dim M_1 + g + 1, g] = [4,2]$ agrees with the Jordan type of $T$-matrix. Furthermore, the RPMHP is $P_{2,0}(q)=q^2+q^4$. Therefore both conjecture \ref{conjn0} and \ref{conj:chavar} hold in this case.

For general $g$, vortex  defects indices of 4d theories give rise to $g$ independent characters $\mathcal{I}_{g, 0}^\text{vortex}(k)$, $k = 0, 2, \ldots, 2g - 2$ \cite{Zheng:2022zkm}, which are all linear combinations of $\eta(\tau)^{2g - 2}E_{2\ell}(\tau)$, $\ell = 0, 1, \ldots, g - 1$. This suggests $g$ ordinary modules with $h = 0, -1, -2, \ldots, g-1$. On the other hand, $\CM^T_{g,0}$ precisely contains $g$ elements with $\mu=0 $ and $\mu = \frac{1}{2},\frac{3}{2},\ldots,g-\frac{3}{2}$. The correspondence between $h$ and $\mu$ again follows \eqref{eq:bijectionn0}.

%Denote $h_d = d - g + 1$ with $d = (g - 1) - \frac{k}{2}$, and we identify $\mathcal{I}_{g,0}^\text{vortex}(k)$ with the fixed points of $\mathcal{M}_{g,0}$,
%\begin{equation}
%L(h_d) \leftrightarrow M_d \ .
%\end{equation}
%The relation between conformal dimensions and critical values of the moment map
%\begin{equation}
%h_d = \mu(M_d) - g +\frac{3}{2} - s( M_d ),
%\end{equation}
%
%When $n=0$, the MLDE has exactly $g$ non-log solutions which is exactly the same number as fixed points. Moreover, the conformal dimension $h_d$ is $-d$ for $0\leq d\leq g-1$, so we would like to propose that the character with conformal dimension $h_d$ corresponds to the fixed point $M_{g-1-d}$ for $0\leq d \leq g-2$ and character with conformal dimension $-(g-1)$ corresponds to $M_0$. Therefore we have the following relation between $h$ and $\mu$
%\begin{equation}
%h=\mu(M)-g+\frac{3}{2}-s(M),
%\end{equation}
%where $s(M)$ is the slope of $M$ modulo $1/2$. Explicitly, $s(M_0)=1/2$ and $s(M_d)=0$. This relation is consistent with the relation in \cite{Fredrickson:2017yka,Fredrickson:2017jcf} as there are no $M_0$ when $g=0$.

The modular $T$-matrix can also be computed explicitly, and its Jordan type is $[3g-2, 3g-4,\ldots,g+2,g]$, which matches  $[\dim M_{g -1} + g + 1, \ldots, \dim M_1 + g + 1,g]$. The total dimension of $\bbV^{\text{mod}}_{g, 0}$ (i.e. the total order of  MDE) is then
\begin{equation}
g+\sum_{d=1}^{g-1}(\dim M_d +g +1) = \sum_{k=0}^{g - 1} (2k + g) =g(2g-1),
\end{equation}
agreeing with equation (37) (where $s=0$)  of \cite{Beem:2021zvt}.

%{\bf Conjecture:} There is a bijection between ordinary modules of unflavored MLDE and fixed points $M_{d,\{0,1,\cdots,1\}}:=M_{d}$. Moreover
%\begin{equation}
%h_M-h_{max} = \mu_M-\mu_{max}|_{\alpha_i=0},
%\end{equation}
%and each $M_{d,\{0,1,\cdots,1\}}$ corresponds t a Jordan block of size $\dim M_{d,\{0,1,\cdots,1\}}+g+1$.

%

\subsection{\texorpdfstring{$g, n>0$ with $n$ even}{}}

%Now we take into account of the non-ordinary modules that are expected to exist due to the $SU(2)^n$ flavor symmetry. Although there is no systematic way of constructing the full spectrum, we can still obtain at least part of modules by studying 4d BPS surface and line defects. Below we compare the prediction from non-local operators with the results from the fixed points. Building on some simple examples presented below, we make the following conjecture concerning the conformal weight of the modules and the moment map on the fixed manifolds.
%
%\begin{conj}Let $h$ (resp. $\Lambda$) be the conformal dimension (resp. weight under $SU(2)^{\otimes n}$) of the highest weight vector, then the formal sum of weights $h$ and $\Lambda$
%\begin{align}
%h+\Lambda= & \ (\mu-\mu_\text{max})|_{\alpha_i\mapsto -\omega^{(i)}}, & \ n ~ & \ \text{even} \ , \\
%h+\Lambda=& \ (\mu-\mu_\text{max})|_{\alpha_i\mapsto \frac{1}{2}+\omega^{(i)}}, & n~ & \ \text{odd} \ .
%\end{align}
%\end{conj}
%

{\bf Example $g=0$, $n=4$:} $V_{0,4}$ is isomorphic to the affine vertex algebra $L_{-2}(D_4)$ whose SMs are classified in \cite{pervse2013note, Arakawa_2016}. There are five SMs with HW $-2\Lambda_0$, $-2\Lambda_1$, $-\Lambda_2$, $-2\Lambda_3$ and $-2\Lambda_4$. To compare with $\CM^T_{0,4}$ we choose the $\fsu(2)^{\otimes 4}$ subalgebra whose simple roots are $\theta$, $\alpha_1$, $\alpha_3$ and $\alpha_4$ of $\fso(8)$ respectively. Results are summarized in table \ref{table:g0n4fps} and confirms the first part of conjecture \ref{conj2}.
\begin{table}[h]
\begin{tabular}{c|c|c||c|c}
$L(\Lambda_i)$ & $h$ & $\fsu(2)^{\otimes 4}$ rep & $M_{d,e}$  & $\mu-\mu_{max}|_{\alpha_i\mapsto -\omega^{(i)}}$ \\ \hline
$-\Lambda_2$ & $-1$ & $[-2,0,0,0]$ & $M_{-3,(0,1,1,1)}$ & $-1-2\omega^{(1)}$ \\ \hline
$-2\Lambda_1$ & $-1$  & $[-2,-2,0,0]$ &$M_{-2,(0,0,1,1)}$  & $-1-2\omega^{(1)}-2\omega^{(2)}$  \\
$-2\Lambda_3$ & $-1$ & $[-2,0,-2,0]$ & $M_{-2,(0,1,0,1)}$  & $-1-2\omega^{(1)}-2\omega^{(3)}$  \\
$-2\Lambda_4$ & $-1$ & $[-2,0,0,-2]$ & $M_{-2,(0,1,1,0)}$  & $-1-2\omega^{(1)}-2\omega^{(4)}$  \\
$-2\Lambda_0$ & $0$ & $[0,0,0,0]$ & $M_{-3,(1,1,1,1)}$  & $0$
\end{tabular}
\caption{\label{table:g0n4fps}$g=0$, $n=4$.}
\end{table}
Using closed form formula of the Schur index and defect indices \cite{Gukov:2008sn,Pan:2021mrw}, the Jordan type of $T$ matrix is $[2,1^4]$ \cite{Zheng:2022zkm}. On the Hitchin side, $M_{-3,(0,1,1,1)}$ has dimension $1$, while the other four FMs have dimension $0$, which also gives $[2,1^4]$ using the second part of conjecture \ref{conj2}. Furthermore the RMHP is $P_{0,4}=q(4+q),$ which confirms conjecture \ref{conj:chavar}.

The vacuum module $L(-2\Lambda_0)$ is the only ordinary module of $V_{0,4}$. Its corresponding FM $M_{-3,(1,1,1,1)}$ is also the FM with largest $\mu$. The Jordan block of $T^\text{ord}$ is $[2]$ which is the same as $\dim M_{-3,(1,1,1,1)}+g+2$. This is an example of corollary \ref{corollary1}.

{\bf Example $(g,n)=(1,2)$:} Schur and defects indices suggest four SMs with $h=0,-1,-1,-1$ \cite{Guo:2023mkn}, whose characters belong to a dim $8$ representation of the modular $SL(2,\bbZ)$. The Jordan type of $T$ is $[3,2^2,1]$. On the Hitchin side, $\CM^T_{1,2}$ has exactly four FMs. The matching is summarized in table \ref{table:g1n2fps}. Plugging dimensions in \eqref{eq:dimJord} yields also $[3,2^2,1]$. The RMHP is
\begin{equation}
P_{1,2}(q)=q(1+2q+q^2).
\end{equation}
\begin{table}[h]
  \begin{tabular}{c|c||c|c|c}
		$h$ & $\fsu(2)^{\otimes 2}$ rep & $M_{d,e}$ & $\dim$  & $\mu-\mu_{max}|_{\alpha_i\mapsto \omega^{(i)} }$\\ \hline
		$-1$ & $[-1,-1]$ & $M_{0}$ & $-$ & $-1-\omega^{(1)}-\omega^{(2)}$\\\hline
		$-1$ & $[-2,-2]$ & $M_{0,(0,0)}$ & $0$ & $-1-2\omega^{(1)}-2\omega^{(2)}$\\ \hline
		$-1$ & $[-2,0]$ & $M_{-1,(0,1)}$ & $1$ & $-1-2\omega^{(1)}$\\ \hline
		$0$ &$[0,0]$ & $M_{-1,(1,1)}$ & $0$ & $0$
  \end{tabular}
  \caption{\label{table:g1n2fps}$g=1$, $n=2$. The dimension of $M_0$ is not used in all conjectures.}
\end{table}

The vacuum module of $V_{1,2}$ which corresponds to $M_{-1,(1,1)}$ is the only OM. The Jordan type of $T^\text{ord}$ is $[3]$ which is the same as $\dim M_{-1,(1,1)}+g+2$.

\subsection{\texorpdfstring{$g, n\geq0$ with $n$ odd}{}}

%{\bf Conjecture:} There is a bijection between non-log solutions of FMLDE and fixed points. Moreover
%\begin{equation}
%h_M=\mu_M-\mu_{max}|_{\alpha_i=0}-s(M),
%\end{equation}
%and each $M_{d,e}$ corresponds to a Jordan block of size $\dim M_{d,e}+g+1$ while $M_0$ corresponds to a Jordan block of size $g$.

% \subsection{\texorpdfstring{$g, n\geq0$ with $n$ odd}{}}

% For theories with odd $n$, we propose the following conjecture relating the conformal weight and highest weight of the simple modules with the moment maps on the Hitchin fixed manifolds.

% \begin{conj}Let $h$ (resp. $\Lambda$) be the conformal dimension (resp. weight under $SU(2)^{\otimes n}$) of the highest weight vector. Then the formal sum
% \begin{equation}
% h+\Lambda=(\mu-\mu_{max})|_{\alpha_i\mapsto \frac{1}{2}+\omega^{(i)}}.
% \end{equation}
% \end{conj}

{\bf Example $g=1$, $n=1$:} $V_{1,1}$ is isomorphic (up to a decoupled $\beta \gamma$ system) to $2d$ small $\CN=4$ superconformal algebra whose SMs are classified in \cite{Adamovic:2014lra}. The matching is summarized in Table \ref{table:g1n1fps}. The indices give rise to a three dimensional representations of the modular $\Gamma^0(2)$, and the Jordan type of $STS$ is $[2,1]$, which is the same as $[\dim M_{0,0}+g+1,g]$. The RMHP is
\begin{equation}
P_{1,1}(q)=q(1+q).
\end{equation}

\begin{table}[h]
\begin{tabular}{c|c||c|c|c}
$h$ &$\fsu(2)$ rep &  $M_{d,e}$ & $\dim$  &   $(\mu-\mu_{max})|_{\alpha\mapsto \frac{1}{2}+\omega}$ \\ \hline
$-1/2$ & $[-1]$   &$M_{0}$ & $-$  &$-1/2-\omega$  \\ \hline
$0$ & $[0]$ & $M_{0,0}$ & $0$  & $0$ 
\end{tabular}
\caption{\label{table:g1n1fps}$g=1$, $n=1$.  The dimension of $M_0$ is not used in all conjectures.}
\end{table}

The vacuum module corresponding to $M_{0,0}$ is the only OM of $V_{1,1}$. The Jordan type of $(STS)^\text{ord}$ is $[2] = [\dim M_{0,0} +g+1]$.
% \begin{table}[h]
% \begin{tabular}{c|c||c|c|c|c|c}
% $\Lambda$ & $h$ & $M_{d,e}$ & $\dim$  & $\mu$ & $\mu-\mu_{max}$ & $(\mu-\mu_{max})|_{\alpha\mapsto \frac{1}{2}+\omega}$ \\ \hline
% $0$ & $0$ & $M_{0,0}$ & $0$ & $\alpha$ & $0$ & $0$ \\ \hline
% $-\omega$ & $-1/2$  &$M_{0}$ & ? & $0$ & $-\alpha$ & $-1/2-\omega$  
% \end{tabular}
% \caption{\label{table:g1n1fps}Fixed points for $g=1$, $n=1$.}
% \end{table}
%There are two fixed points in total. One is $M_{0,0}$ with $\mu_{0,0} = \alpha$ and $\dim M_{0,0}=0$. The other one is $M_0$ which is diffeomorphic to the moduli space of stable rank-2 bundles  over $\Sigma_{1,1}$.
%
%The FMLDEs are {\color{red} MLDE and solutions}.
%
%We propose that $M_{0,0}$ corresponds to the vacuum module while $M_0$ corresponds to the other simple module from solutions of FMLDE. 
%Again we have
%\begin{equation}
%	h_M=\mu_M-\mu_{max}|_{\alpha=0}-s(M),
%\end{equation}
%and the alpha dependent part of $\mu$ becomes the finite weight of the corresponding VOA module after setting $\alpha=\omega$. FMLDE implies that $\bbV_{1,1}$ is composed of $2$ simple modules and $1$ log module with the Jordan type of $T$ being $[2,1]$, which is the same as $\dim M_{0,0}+g+1$ and $g$. 

{\bf Example $g=2$, $n=1$:} $V_{2,1}$ has four SMs whose character belongs to a $14$ dimensional representation of the modular group. The matching is shown in Table \ref{table:g2n1fps}.  The Jordan type of $STS$ is $[5,4,3,2]$. The RMHP is
\begin{equation}
q^2(1+q+q^2+q^3).
\end{equation}
\begin{table}[h]
\begin{tabular}{c|c||c|c|c}
$h$ & $\fsu(2)$ rep & $M_{d,e}$ & $\dim$  & $(\mu-\mu_{max})|_{\alpha\mapsto \frac{1}{2}+\omega}$\\ \hline
$-3/2$ & $[-1]$ & $M_{0}$ & $-$ & $-3/2-\omega$ \\ \hline
 $-1$ & $[-2]$ & $M_{0,1}$ & $1$ & $-1-2\omega$ \\ \hline
$-1$ & $[0]$ & $M_{0,0}$ & $2$ & $-1$ \\ \hline
$0$ & $[0]$ & $M_{1,0}$ & $0$ & $0$
\end{tabular}
\caption{\label{table:g2n1fps}$g=2$, $n=1$. The dimension of $M_0$ is not used in all conjectures.}
\end{table}

There are two OMs, corresponding to $M_{1,0}$ and $M_{0,0}$. The Jordan type of $(STS)^\text{ord}$ is $[5,3] = [\dim M_{0,0} + g + 1, \dim M_{1,0} + g + 1]$ as expected.

\subsection{Ordinary modules of \texorpdfstring{$V_{g,n > 0}$}{Vgn}}

We can systematically construct the characters of OMs of $V_{g,n}$ from the (defect) Schur indices and modular transformations \cite{Pan:2024bne}, providing additional examples to support corollary \ref{corollary1} besides those mentioned earlier. We summarize the results in table \ref{table:ordeven}. The difference $\mu - \mu_\text{max}$ reproduces the expected conformal weight $h$ of the corresponding modules, while  dimensions of FMs (denoted as  subscripts of $\mu - \mu_\text{max}$) give the Jordan type of $T^\text{ord}$ or $(STS)^\text{ord}$ via the formula in corollary \ref{corollary1}. The total dimension of $\bbV^{\text{ord}}_{g, n}$ (i.e. the total order of  MDE) matches equation (37) (at $s_\text{there}=n_\text{here}$)  of \cite{Beem:2021zvt}.
\begin{table}[h]
\begin{tabular}{c||c|c||c}
$g,n$ & $h$ & $Jordan$ & $\mu-\mu_{max}$ \\ \hline
%$0,4$ & $0$ & $[2]$ & $0_1$ \\ \hline
$0,6$ & $0,-1$ &  $[4,2]$ & $-1_2, 0_0$ \\ \hline
$0,8$ & $0,-1,-2$ &  $[6,4,2]$ & $-2_4, -1_2, 0_0$ \\ \hline
%$1,2$ & $0$ &  $[3]$ & $0_1$ \\ \hline
$1,4$ & $0,-1$ &  $[5,3]$ & $-1_2, 0_0$ \\ \hline
$1,6$ & $0,-1,-2$ &  $[7,5,3]$ & $-2_4,-1_2,0_0$ \\ \hline
$2,2$ & $0,-1$ &  $[6,4]$ & $-1_2,0_0$ \\ \hline
$2,4$ & $0,-1,-2$ &  $[8,6,4]$ & $-2_4,-1_2,0_0$ \\ \hline \hline
$1,3$ & $0,-1$ &  $[4,2]$ & $-1_2, 0_0$ \\ \hline
$2,1$ & $0,-1$ &  $[5,3]$ & $-1_2, 0_0$ 
%$2,1$ & $0,-1$ &  $[5,3]$ & $-1_2, 0_0$ \\ \hline
\end{tabular}
\caption{\label{table:ordeven}Matching OMs and FMs. In the last column, the subscript of each $\mu-\mu_{max}$ is the dimension of the corresponding FM. }
\end{table}

\section{Generalizations to higher rank}
\label{sec:higherrank}

We expect the 4d mirror symmetry for higher rank class-$\cal S$ theories. Notable examples include the Minahan-Nemeschansky $E_6$, $E_7$, $E_8$ theories and our conjecture holds for all these cases \cite{Arakawa_2016, Shan:2023xtw}.

In higher rank cases, it is much easier to compute RMHP of the character variety. For example,  consider the $SU(3)$ $\CN=2^\ast$ theory, which is the compactification of $6d$ $A_2$ SCFT on a torus with one minimal puncture. The corresponding VOA $V^{\fsu(3)}_{1,1}$ is constructed in \cite{Beem:2013sza, Bonetti:2018fqz, Arakawa:2023cki}. The RMHP from character variety is
\begin{equation}
q(1+q+q^2).
\end{equation}
Following our conjecture, we predict there to be $3$ simple modules whose characters are in a 6 dimension representation of the modular group $\Gamma^0(2)$. This prediction is compatible with the Wilson line index computation in 4d, where the candidate characters are given by  indice $\mathcal{I}_{\mathcal{N} = 4}(b)$, $\frac{\vartheta_4(\mathfrak{b})}{\vartheta_4(3 \mathfrak{b})}$ and $\frac{\vartheta_4(\mathfrak{b})}{\vartheta_4(3 \mathfrak{b})}(E_1 \big[\substack{-1 \\ b}\big] + E_1 \big[\substack{-1 \\ b^2 q^{\frac{1}{2}}}\big])$, generating a six-dimensional representation as expected.

\section{Outlook}
\label{sec:outlook}

In this letter we systematically study the correspondence between the representation theory of $V_{g,n}$ and geometry of $\CM_{g,n}$ for $A_1$ class-{\cal S} theories. Recently in 3d $\CN=4$ theory, the relation between chiral quantization of HB and coordinate ring of CB is discussed \cite{Costello:2018swh, Beem:2023dub}. It would be interesting to discuss connections between this 3d correspondence and our results, and might provdie in sights on the physical reason behind the 4d mirror symmetry. 
It would also be interesting to see whether our story has any connection with free field realization of $V_{g,n}$ \cite{Beem:2022mde, Beem:2022vfz, Beem:2024fom}.

Another question is to find the CB counterpart of the log modules. When $g=0$, each FM is isomorphic $\bbP^d$, hence $\dim H^{\ast}(M_{d,e}) = d+1$, and we propose that both SMs and log modules are in bijection with elements of the cohomology ring of $\CM_{0,n}$. When $g>0$ the extra $g$ in the formula hints that extra log modules might connected to elements in $H^1$ part of the cohomology.

%Another related question is how the modular group appears in the Coulomb branch side. In general AD theory, the action of modular group was realized by the automorphism of double affine Hecke algebra acting on the cohomology of fixed points. In class-S cases, the modular group action may also realized by certain action on the cohomology of fixed points as well. Identifying such action should also help us answer the previous question.
%

\begin{acknowledgments}
The authors would like to thank Drazen Adamovic, Tomoyuki Arakawa, Penghui Li, Peng Shan and Dan Xie for helpful discussions. W.Y. is supported by  National Key R\&D Program of China (Grant 2022ZD0117000).
 The work of Y.P. is supported by the National Natural Science Foundation of China (NSFC) under Grant No. 11905301. \end{acknowledgments}

\bibliography{ref}

\end{document}